\documentclass{pasa}%

\usepackage{graphicx}
\usepackage[natbib, style=apa, mincitenames=3, maxcitenames=3]{biblatex}
\title[Improved MWA SDA]{Improved Sensitivity for Space Domain Awareness Observations with the Murchison Widefield Array}

\author[Prabu et al.]{Prabu, S.$^{1,4}$, Hancock, P.$^2$, Zhang, X.$^{3,5}$, Tingay, S.J.$^{1}$, Hodgson, T.$^{1,2}$, 
Crosse, B.$^1$, and Johnston-Hollitt, M.$^2$
\affil{$^1$International Centre for Radio Astronomy Research, Curtin University, Bentley, WA 6102, Australia}%
\affil{$^2$Curtin Institute for Computation, Curtin University, GPO Box U1987, Perth, 6845, WA, Australia}%
\affil{$^3$CSIRO Space and Astronomy, 26 Dick Perry Avenue, Kensington, WA 6151, Australia}
\affil{$^4$CSIRO Space and Astronomy, Corner Vimiera \& Pembroke Roads, Marsfield, NSW 2122, Australia}
\affil{$^5$Shanghai Astronomical Observatory, 80 Nandan Road, Xuhui District, Shanghai 200030, China}
}%

\jid{PASA}
\doi{10.1017/pas.\the\year.xxx}
\jyear{\the\year}

\usepackage{aas_macros}
\usepackage{hyperref} 
\hypersetup{colorlinks,citecolor=blue,linkcolor=blue,urlcolor=blue}

\begin{document}

\begin{frontmatter}
\maketitle

\begin{abstract}
Our previously reported survey of the Low Earth Orbit (LEO) environment using the Murchison Widefield Array (MWA) detected over 70 unique Resident Space Objects (RSOs) over multiple passes, from 20 hours of observations in passive radar mode. In this paper, we extend this work by demonstrating two methods that improve the detection sensitivity of the system. The first method, called shift-stacking, increases the statistical significance of faint RSO signals through the spatially coherent integration of the reflected signal along the RSO's trajectory across the sky. This method was tested on the observations used during our previous blind survey, and we obtained a $75\%$ increase in the total number of detections. The second method re-focuses the MWA to the near-field RSO's position (post-observation), by applying a complex phase correction to each visibility to account for the curved wave-front. The method was tested successfully on an MWA extended array observation of an ISS pass. However, the method is currently limited by signal de-coherence on the long-baselines (due to the hardware constraints of the current correlator).  We discuss the sensitivity improvement for RSO detections we expect from the MWA Phase 3 correlator upgrade.  We conclude the paper by briefly commenting on future dedicated Space Domain Awareness (SDA) systems that will  incorporate MWA technologies.
\end{abstract}

\end{frontmatter}

\section{INTRODUCTION }
\label{sec:intro}
The onset of the Kessler Syndrome \citep{1978JGR....83.2637K}, a cascading collision event scenario in the near-Earth environment, can be delayed or prevented by preforming space surveillance. The current rapid increase in the number of Resident Space Objects (RSOs) in the Low Earth Orbit (LEO) demands the development of multiple Space Domain Awareness (SDA) sensors, that together contribute towards the global SDA effort. Most of the current SDA activities are undertaken by the Space Surveillance Network (SSN) \citep{miller2007new} based in the US and its European equivalent operated by the European Space Agency (ESA) \citep{Bobrinsky2010}. The majority of the existing SDA facilities consist of either optical sensors that perform space surveillance during twilight or active coherent radars that perform searches for RSO reflections (we direct the reader to \cite{2021ApSci..11.1364M} for a detailed summary of current SDA radar systems and how the non-coherent MWA radar system employed in this work is novel compared to existing solutions). Both of these types of systems often have a small Field of View (FOV), thus limiting their capability to perform simultaneous detections.

In the recent past, there has been a growing interest in decentralising SDA information across different countries in order to prevent ``bias of information'' \citep{lal2018global}. We aim to address these issues by using an already existing wide FOV radio telescope, the Murchison Widefield Array (MWA) \citep{Tingay2013TheFrequencies} \citep{2018PASA...35...33W}, as a passive radar to perform space surveillance by searching for RSO reflections of terrestrial FM transmissions. The MWA is an international effort by different countries and its data are publicly available\footnote{the observations are publicly made available after its 18 months of proprietary period, in order to protect the interest of the science groups performing the observations.}. In this paper, we build upon our previous SDA work with the MWA, and demonstrate two different techniques that provide more sensitive RSO signal detections. 

In our previous survey of the LEO environment using the MWA \citep{prabu_survey}, we demonstrated blind detection techniques for detecting RSOs, meteors, and aircraft using MWA observations. Whilst blind searches are a good method for detecting lost/new RSOs, in this paper we perform more sensitive detections of faint RSO signals by using prior knowledge of their trajectory to integrate the signal across the sky.

The first method that we explore in this work is called ``shift-stacking''. The shift-stacked RSO search method uses phase-tracked difference images to perform detections by integrating the signal along the predicted trajectory of the pass. Although shift-stacking has been previously described in the literature to search for new Kuiper belt objects, trans-Neptunian objects \citep{2004AJ....128.1364B} such as ``Planet Nine'' \citep{2020PSJ.....1...81R}, and solar-system satellites \citep{2016AJ....151..162B} by trial and error iterations performed in a multi-dimensional parameter space, we adapt this method to perform  searches for weak RSO signals in MWA data. A similar method has been also previously explored by \cite{tagawa2016orbital} for optical SDA sensors.


The second method that is demonstrated in this paper aims to improve the detection of RSO signals by re-focusing the interferometer (the MWA) to the predicted near-field RSO position. Standard interferometer theory assumes the observed source to be in the far-field of the instrument, thus deriving a 2D Fourier relationship between the re-constructed sky image and the measured visibilities. However, due to the near-field nature of the LEO RSOs, the longer baselines of the MWA see a curved wave-front rather than a planar wave-front. Hence, when imaged (without accounting for the curvature), the satellite signal appears de-focused, resulting in reduced signal to noise. In this work, we demonstrate a near-field RSO search performed using the MWA and also address its current limitations due to hardware constraints.

This paper is structured as follows. In Section \ref{background} we briefly describe the MWA and the SDA techniques. In Section \ref{dataandmethods} we describe the observations used and the data processing methods employed. The results of the analysis performed are provided in Section \ref{resultsanddiscussion}, followed by a brief discussion. We draw our conclusions in Section \ref{conclusion}.

\subsection{BACKGROUND}
\label{background}
Most of the current SDA sensors are concentrated in the northern hemisphere.  The development of SDA capability using the limited land in the Southern Hemisphere is therefore of international importance. Hence, since the MWA has been shown in the past to be capable of performing SDA observations, we continue our effort to develop further the sensitivity of the MWA SDA system. The development of more sensitive detection methods was also motivated by our participation in Space Fest\footnote{\url{https://www.airforce.gov.au/our-mission/spacefest-edge}} 2020, an Australian Airforce sponsored civilian event, when different SDA groups from around Australia came together to demonstrate their SDA capabilities. Based in part on the demonstrations performed during Space Fest, future dedicated SDA facilities may be constructed.

The MWA \citep{Tingay2013TheFrequencies} is a low-frequency radio interferometer located at the radio quiet Murchison Radio-astronomy Observatory (MRO), Western Australia. The MWA is capable of observing the southern radio sky between $70 - 300$\, MHz using $128$ tiles\footnote{a $4\times 4$ array of duel-polarised bow tie antennas. Images of MWA tiles can be found in \url{https://www.mwatelescope.org/multimedia/images}}, with an instantaneous bandwidth of $30.72$\,MHz. The MWA underwent an upgrade to Phase $2$ during 2016 \citep{2018PASA...35...33W}, using the same hardware as the Phase $1$ array, but with an additional $128$ tiles deployed. The MWA Phase 2 array is periodically reconfigured between the \textit{extended} configuration and the \textit{compact} configuration, each of which contains $128$ tiles \citep{2018PASA...35...33W}. The extended configuration has baseline lengths up to $5.3$\,km and has a higher angular resolution\footnote{compared to MWA Phase 1 configuration and the Phase 2 compact configuration}, while most of the compact configuration baselines are under $200$\,m to provide increased sensitivity to extended emission. 

The MWA previously demonstrated SDA by performing coherent SDA radar detections \citep{7944483}\citep{8835821}\citep{hennessy2020orbit} using the MWA's Voltage Capture System \citep{2015PASA...32....5T}, as well as non-coherent SDA radar detections \citep{Tingay2013OnFeasibility}\citep{Zhang2018LimitsMWA}\citep{prabu_dev}\citep{ prabu_survey} using the MWA's standard operation mode (as an aperture synthesis array). By using the line of sight FM transmission as the reference signal, the coherent radar method uses matched filtering to identify satellite reflections.  The non-coherent radar method performs source finding on difference images to search for satellite signals. In this paper, we continue to develop the non-coherent method, and the reader is referred to the aforementioned  papers for more information about the MWA's coherent SDA system. 

The non-coherent MWA SDA system performs RSO detections by searching for time-varying signals in fine-channel ($40$\,kHz wide) difference images. \cite{Zhang2018LimitsMWA} demonstrated difference imaging to be an effective technique to search for RSOs in MWA observations, as it subtracts the static background sources along with its side-lobe confusion noise and is only limited by the thermal noise of the instrument. RSO signals in difference images appear as a streak with a positive head and a negative tail, as shown in Figure \ref{Fig1shiftstack} and Figure \ref{Fig7near-field}.

\section{DATA AND METHODS}
\label{dataandmethods}
\subsection{Shift-Stacking}
\label{shift-stacking}

\begin{figure*}[h!]
\begin{center}
\includegraphics[width=\linewidth,keepaspectratio]{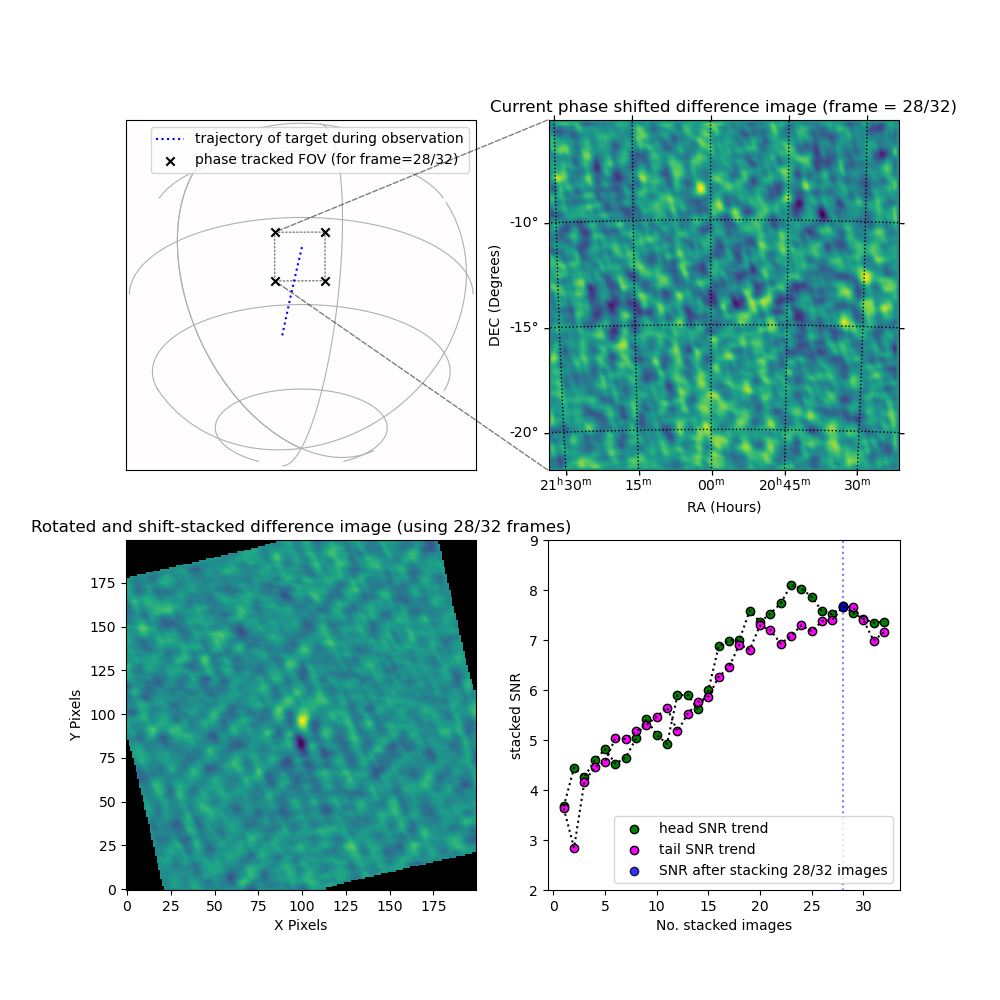}
\caption[Demonstration of shift-stacking.]{Demonstration of detection performed using the shift-stacking method. The top left-panel is the entire horizon visible to the MWA during the observation. The predicted trajectory of the satellite (blue dotted line) and the phase tracked FOV of the satellite for frame $N=28$ (black crosses) are shown. The insert panel (top-right) shows the phase-tracked fine channel difference image for the considered time-step/frame. The bottom-left panel is the rotated shift-stacked fine channel difference image of the RSO signal after stacking N frames (in this case $N=28$). An animation of this figure is available at \url{https://www.youtube.com/watch?v=_dbEW61RnFk}. The bottom-right panel shows the SNR of the RSO signal increasing with the number of stacked frames.}
\label{Fig1shiftstack}
\end{center}
\end{figure*}

We test the shift-stacking method using the $20$ hours of Phase 2 compact configuration observations used in our previous blind survey \citep{prabu_survey}. For the MWA, the GPS time of the observation is used as the observation ID, and observations can be obtained from the All-Sky Virtual Observatory (ASVO)\footnote{\url{https://asvo.mwatelescope.org/}}. Re-using the observations in this work allows us to compare the performance of the previously demonstrated blind detection method and the shift-stacking detection method developed here. The visibilities are calibrated (using the {\tt calibrate} tool developed in \cite{2016MNRAS.458.1057O}) and converted to {\tt CASA} measurement set format \citep{2007ASPC..376..127M} as explained in \cite{prabu_survey}, and the fine frequency channel images are created using {\tt WSClean} \citep{offringa-wsclean-2014}\citep{offringa-wsclean-2017}. 

The shift-stacking method seeks to detect RSOs by performing a spatially coherent averaging of the signals along the predicted trajectory. An example of shift-stacking for the object PICOSAT-9 (NORAD ID $26930$) during the observation ID $1157400472$ is demonstrated in Figure \ref{Fig1shiftstack}. The top-left panel shows the predicted trajectory of the RSO above the visible horizon and the phase-tracked field-of-view (FOV) of an individual shift-stack frame. Due to the large FOV of the MWA (approx. $1300$ degree$^{2}$), the curvature of the RSO pass is often resolved (i.e, the apparent direction of the RSO motion changes during the pass) and  stacking without any orientation correction would result in signal smearing. Hence we rotate the individual frames (using {\tt scipy.ndimage.rotate}\footnote{\url{https://docs.scipy.org/doc/scipy/reference/generated/scipy.ndimage.rotate.html}}) to align the signal in the vertical direction (arbitrarily chosen) prior to stacking. The bottom-left panel of Figure \ref{Fig1shiftstack}
 shows the rotated shift-stacked difference image (after performing an inverse-noise weighted stacking of $N$ frames) for one fine frequency channel.

All LEO RSOs that were predicted to pass through the MWA's primary beam during the observations were identified\footnote{by performing an API query using {\tt spacetracktool.SpaceTrackClient.tle\_publish\_query} for the epoch. \url{https://pypi.org/project/spacetracktool/}}. For every identified object, we search for the satellite signal at every $40$\,kHz fine frequency channel using the shift-stacking pipeline\footnote{ the bash pipeline can be found in \url{https://github.com/StevePrabu/Space-Fest/blob/master/bin/phaseTrack.sh} and the python scripts used can be found in \url{https://github.com/StevePrabu/PawseyPathFiles}} (resulting in a stacked image cube). Prior to stacking, the phase centre of each of the individual frames is centred (using the {\tt chgcentre} tool developed by Andre Offringa) at the predicted location of the object during the epoch, thus enabling spatially coherent stacking of the faint RSO signal along the RSO's trajectory. 

All $6 \sigma$ detections that appeared as the expected streak signal were identified from the shift-stacked image cube (the third dimension being frequency). We vet the detections to obtain a preliminary list\footnote{note that we sort them further to remove false positives in Section \ref{resultsanddiscussion}} of candidate detections using the criterion that the detection must appear as a spatially coherent $6 \sigma$ streak in more than one fine frequency channel. This helps get rid of aliased side-lobes of other bright events that may spatially coincide with the object of interest and this is also the selection  criterion used during our previous blind survey. The list of $164$ detected candidates are further investigated in Section \ref{resultsanddiscussion}.

\subsection{Near-Field Re-Focusing}
\begin{figure*}[h!]
\begin{center}
\includegraphics[width=0.7\linewidth,keepaspectratio]{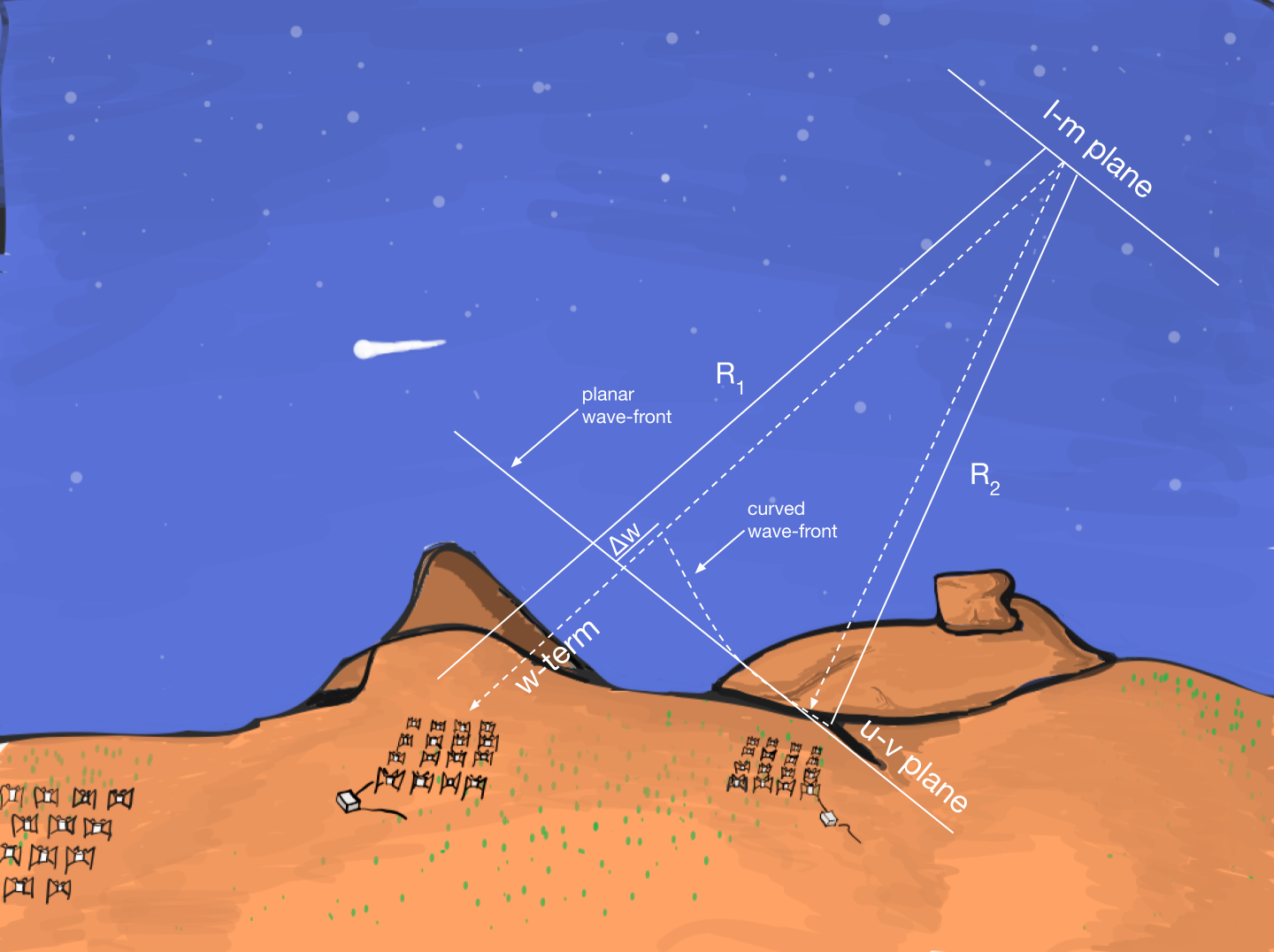}
\caption[The near-field curvature as seen by a baseline pair.]{The near-field curvature as seen by a baseline-pair.  $\Delta w$ is the delay due to the curved wave-front and $R_{i}$ is the distance between the RSO and $tile_{i}$.}
\label{Fig2near-field}
\end{center}
\end{figure*}

The RSO reflected wavefronts appear curved when observed using the long MWA baselines, resulting in de-focused (reduced SNR) signals. This apparent deviation ($\Delta w$) from a planar wave-front  as seen by a ``long'' baseline is shown in Figure \ref{Fig2near-field}. Since the Phase $2$ compact configuration array has the majority of its baselines shorter than $200$\,m, LEO RSOs would appear in the far-field of the instrument and therefore no re-focusing is required. We therefore focus our efforts on the Phase $2$ extended array (observation ID $1290483224$ that had the ISS pass through the primary beam) which has predominantly long baselines that sees the RSO in the near-field. The ISS observation was calibrated using amplitude and phase solutions obtained for the observation ID $1290513616$ (infield GLEAM sources \citep{2017MNRAS.464.1146H} were used as the model for the sky based calibration performed, followed by the application of self-calibration). The two observations were approximately $8$ hours apart and MWA calibration solutions are generally valid within a day.

Prior knowledge of the RSO's and MWA tile's Geocentric Cartesian coordinates (e.g, $X_{rso}$, $Y_{rso}$, $Z_{rso}$ and  $X_{tile1}$, $Y_{tile1}$, $Z_{tile1}$) are used to calculate the actual deviation/delay from a planar wave-front ($\Delta w$ in Figure \ref{Fig2near-field}) as seen by each baseline (calculated independently for every time-step within the observation). The calculated delay $\Delta w$ is applied as a phase offset to re-focus the array to the desired near-field location. For a baseline between $tile_{1}$ and $tile_{2}$, $\Delta w_{1-2}$ is calculated using Equation \ref{E1}

\begin{equation}
    \begin{array}{l}
        R_{1}^2 =(X_{rso} - X_{tile1})^{2} + (Y_{rso} - Y_{tile1})^{2}  \\
        \qquad \qquad \qquad \quad + (Z_{rso} - Z_{tile1})^{2} \\
        R_{2}^2 =(X_{rso} - X_{tile2})^{2} + (Y_{rso} - Y_{tile2})^{2}\\
        \qquad \qquad \qquad \quad + (Z_{rso} - Z_{tile2})^{2} \\
          \Delta w_{1-2} = (R_{1} - R_{2}) - w_{1-2} 
         
    \end{array}
    \label{E1}
\end{equation}

where $R_{1}$($R_{2}$) is the distance in metres between the RSO and 
 $tile_{1}$($tile_{2}$). $\Delta w_{1-2}$ is the w-term associated with a planar wave-front as seen by the baseline. The calculated delay is applied to the measured visibility as a complex phase offset using Equation \ref{E2} (adapted from  \cite{marr2015fundamentals}),
 
 \begin{equation}
         \Delta phase_{1-2} = \exp^{i 2\pi \frac{\Delta w_{1-2}}{\lambda}},
    \label{E2}
\end{equation}

where $\lambda$ is the wavelength of the fine frequency channel. The correction is applied to the visibilities set prior to imaging using our python tool {\tt LEOVision}\footnote{\url{https://github.com/StevePrabu/LEOVision}}.  Results are discussed in Section \ref{resultsanddiscussion}.

\section{RESULTS AND DISCUSSION}
\label{resultsanddiscussion}
\subsection{Shift-Stacking Candidates}
\begin{figure}[h!]
\begin{center}
\includegraphics[width=\linewidth,keepaspectratio]{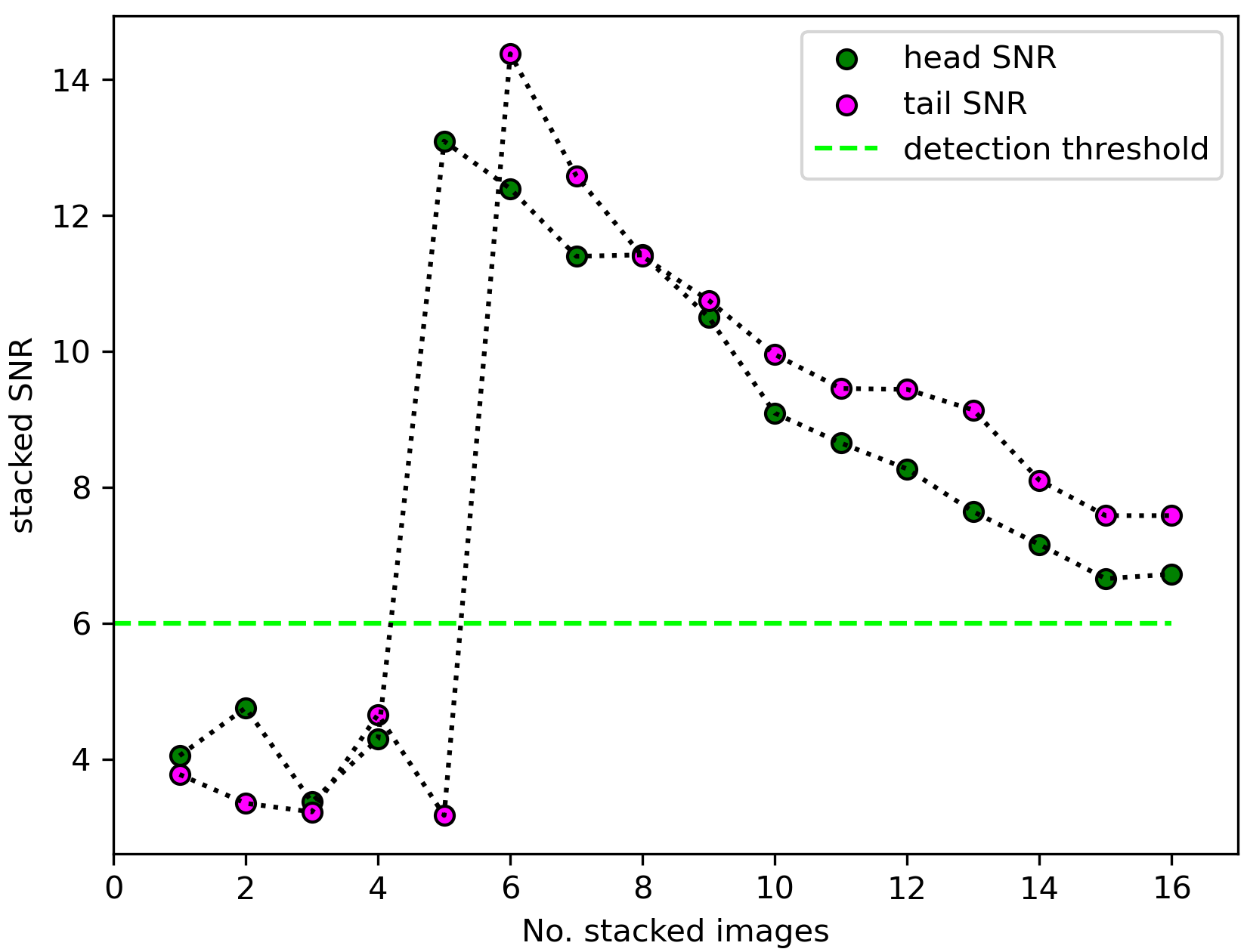}
\caption[An example of a flaring event detected using shift-stacking.]{An example of a flaring event detected using the shift-stack pipeline. Figure shows the SNR of the RSO signal reducing with the number of stacked frames.}
\label{Fig3flaringEvent}
\end{center}
\end{figure}

Through manual inspection of the $164$ candidates obtained from the shift-stack search performed in Section \ref{shift-stacking}, we identify two different types of detections, namely, flaring events (64) and steady reflection events (100). Flaring events often have a single frame with a very high SNR signal, where stacking reduces the SNR of the detection (due to a lack of spatially coherent signals in other frames). An example of a flaring event is shown in Figure \ref{Fig3flaringEvent}. On the contrary, the SNR of steady reflection events only increases with stacking and an example is shown in the bottom panels of Figure \ref{Fig1shiftstack}.

The two different populations of candidates are further analysed based on their  offset from the phase centre (predicted location of the object using previously estimated orbital parameters) of the shift-stacked image, and the frequency channel they are detected in. While the frequency channel investigation helps determine if the event is due to an FM reflection from RSOs, or bad difference imaging (imperfect background source subtraction due to time varying sky signal or instrument response), the apparent offset helps determine if the detected signals are likely to be from the RSO of interest.

\begin{figure}[h!]
\begin{center}
\includegraphics[width=\linewidth,keepaspectratio]{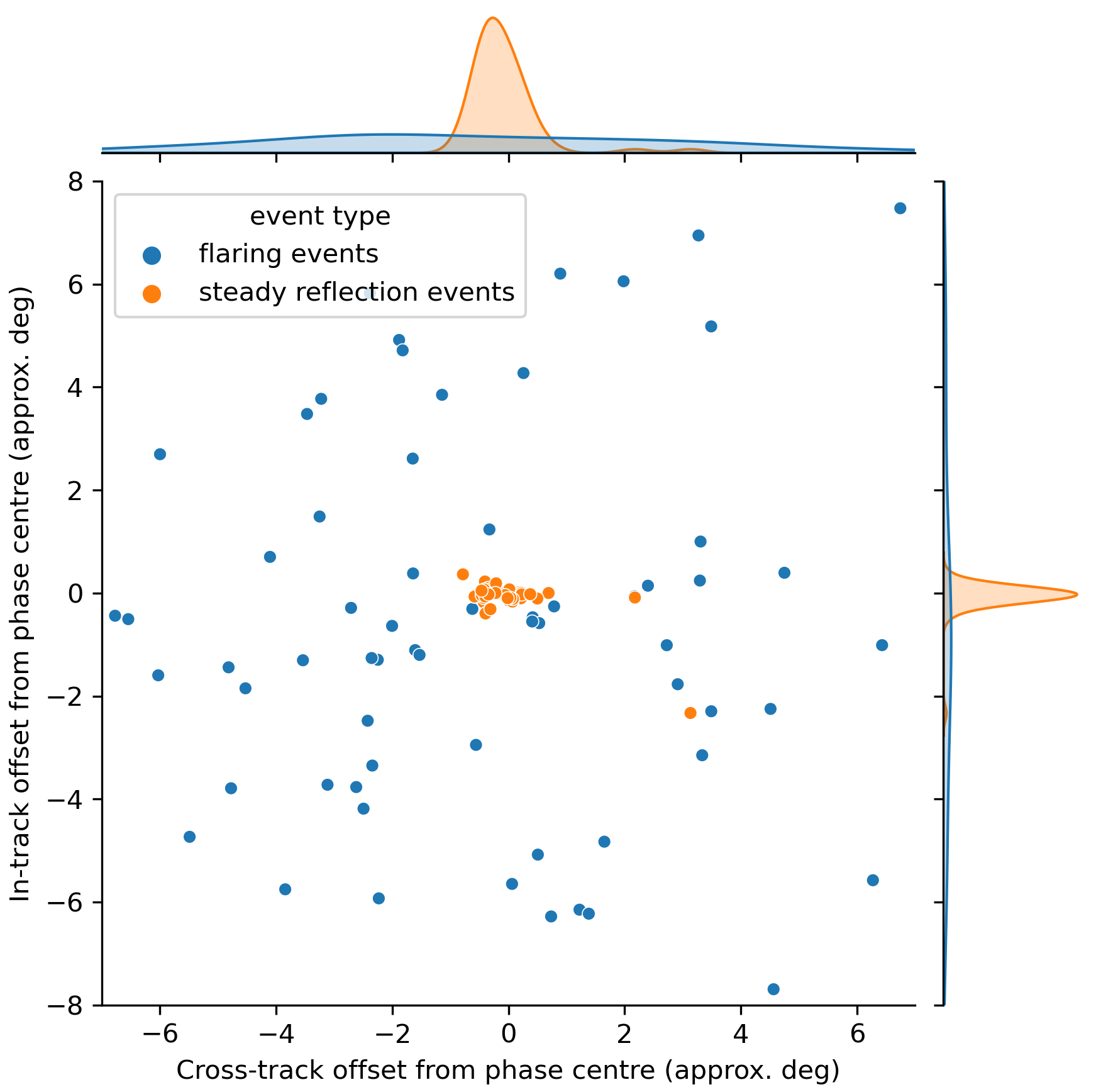}
\caption[Apparent centroid distribution for flaring and steady reflection events.]{The apparent centroid offset distribution for steady reflection and flaring events. Most of the steady reflection events are within a degree of the predicted location, while the flaring events have almost a uniform distribution within the FOV and tapers off towards the edge due to rotating the frames prior to stacking. }
\label{Fig4centroid}
\end{center}
\end{figure}

Figure \ref{Fig4centroid} shows the centroid offset (projected into in-track and cross-track directions) for all the flaring and steady reflection events, and we see very different behaviour between the two populations. The majority of the steady reflection events appear within a degree of the phase centre (expected location of the object), while the flaring events have an almost uniform distribution of offsets before tapering off near the edge of the shift-stacked FOV.
This tapering is due to image rotation resulting in many blank pixels near the edge as  seen in the bottom-left panel of Figure \ref{Fig1shiftstack}.

As most of the steady reflection events are closer to the predicted locations than the flaring events, they are more likely to be from the RSO of interest, and hence we classify them as our final list of candidate detections. We detect almost all the RSOs detected during our blind survey \citep{prabu_survey}, except a few RSOs where a few noisy frames reduced the stacked SNR of the RSO signal below the detection threshold used in Section \ref{shift-stacking}). We also detect many more new RSOs not previously detected. Using shift-stacking, we improve the total number of RSO detection candidates from our $20$ hours of observation by $75\%$\footnote{During the blind survey $73$ events were detected inside the primary beam (along with $7$ detections of large objects outside the primary beam). Using shift-stack search, we obtain $55$ new events inside the primary beam.}. The list of new detections (excluding the events already detected during the blind survey in \cite{prabu_survey}) is given in Table \ref{tab1}. Note that for the MWA, primary beam correction is applied in the image domain, and creating the corresponding phase tracked primary beam models for every time-step is a very computationally intensive task, as the individual shift-stack frames were not primary beam corrected. Hence, Table \ref{tab1} provides only the apparent peak flux density of the detections and can be treated as a lower limit on the actual flux density of the detection. The in-track and cross-track offsets measured for every steady reflection event (shown in Figure    \ref{Fig4centroid}) can also be used for performing orbital element updates for the RSOs.

\begin{figure}[h!]
\begin{center}
\includegraphics[width=\linewidth,keepaspectratio]{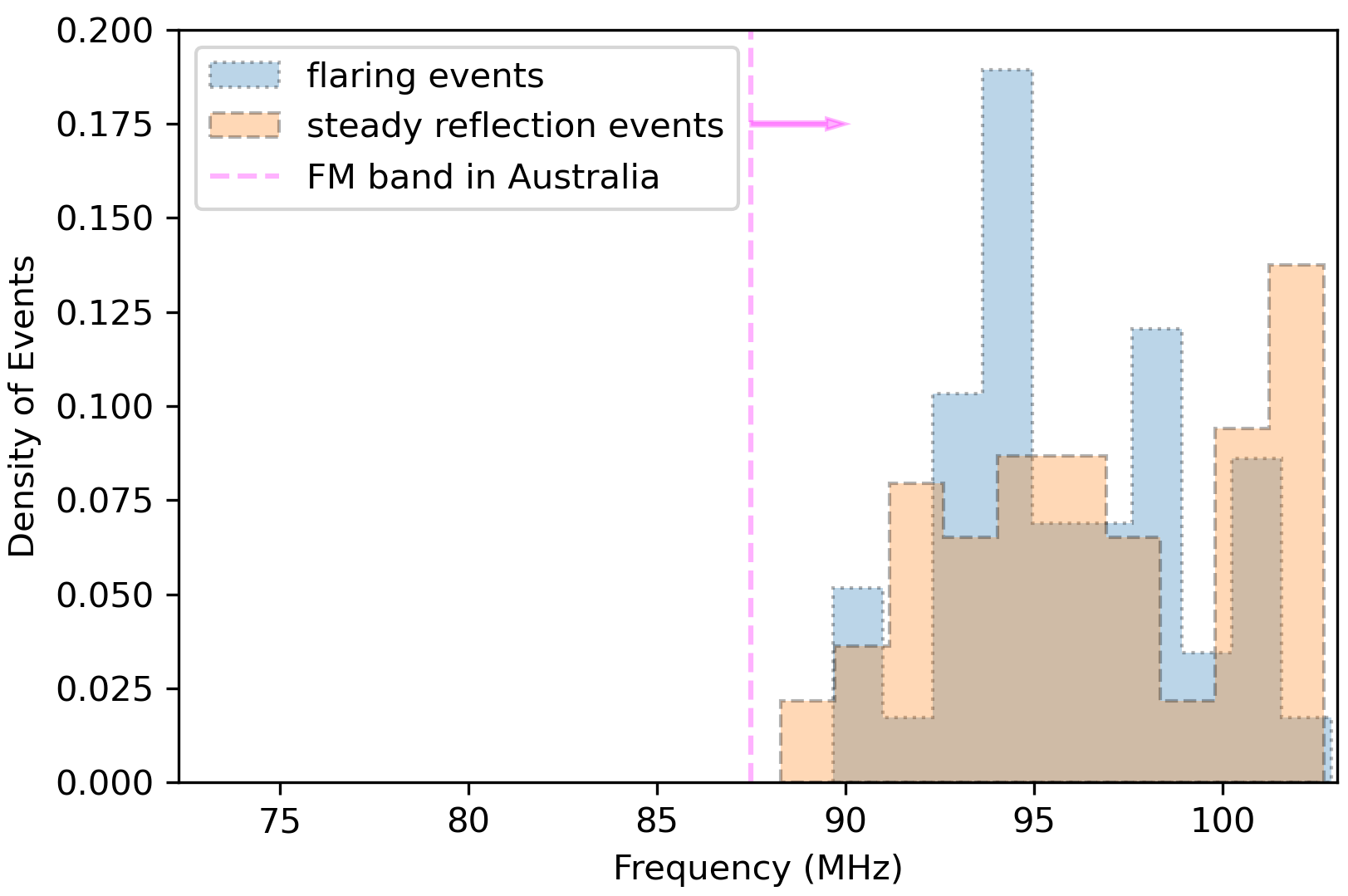}
\caption[Frequency distribution for flaring and steady reflection events.]{The density distribution of channels in which the flaring and steady reflection events were detected. The two brightest channels at which the event was detected spatially coherent (channels identified during the candidate vetting process in Section \ref{shift-stacking}) was used to create the plot. We see that all the steady reflection events are confined within the FM band (as expected), and all of the flaring events were also detected within the FM band, implying that many of the flaring events could be FM reflections from other real events such as meteors trails.}
\label{Fig5frequency}
\end{center}
\end{figure}

The frequency of the observations used in the shift-stack analysis span  $72.335 - 103.015$\,MHz, while the FM band only overlaps with the second half of the observation band (i.e, above $88$\,MHz). Figure \ref{Fig5frequency} shows the frequency distribution of all the detected steady reflection and flaring events. From Figure \ref{Fig5frequency} we see that all the steady reflection events are confined within the FM band (as expected from FM reflecting RSO) and all of the flaring events are also confined within the FM band. Hence, it is very likely that the flaring events, though not the RSOs searched for, are real reflection events (and not due to noise). Many ($8.5$ hours) of the observations used coincided with the Geminids meteor shower and it is likely that many of the flaring events are associated with FM reflections from ionised meteor trails. The occurrence of these FM reflecting flaring events are consistent with a recent study of the RFI environment at the MRO by \cite{2020PASA...37...39T}.

\begin{figure*}[h!]
\begin{center}
\includegraphics[width=\linewidth,keepaspectratio]{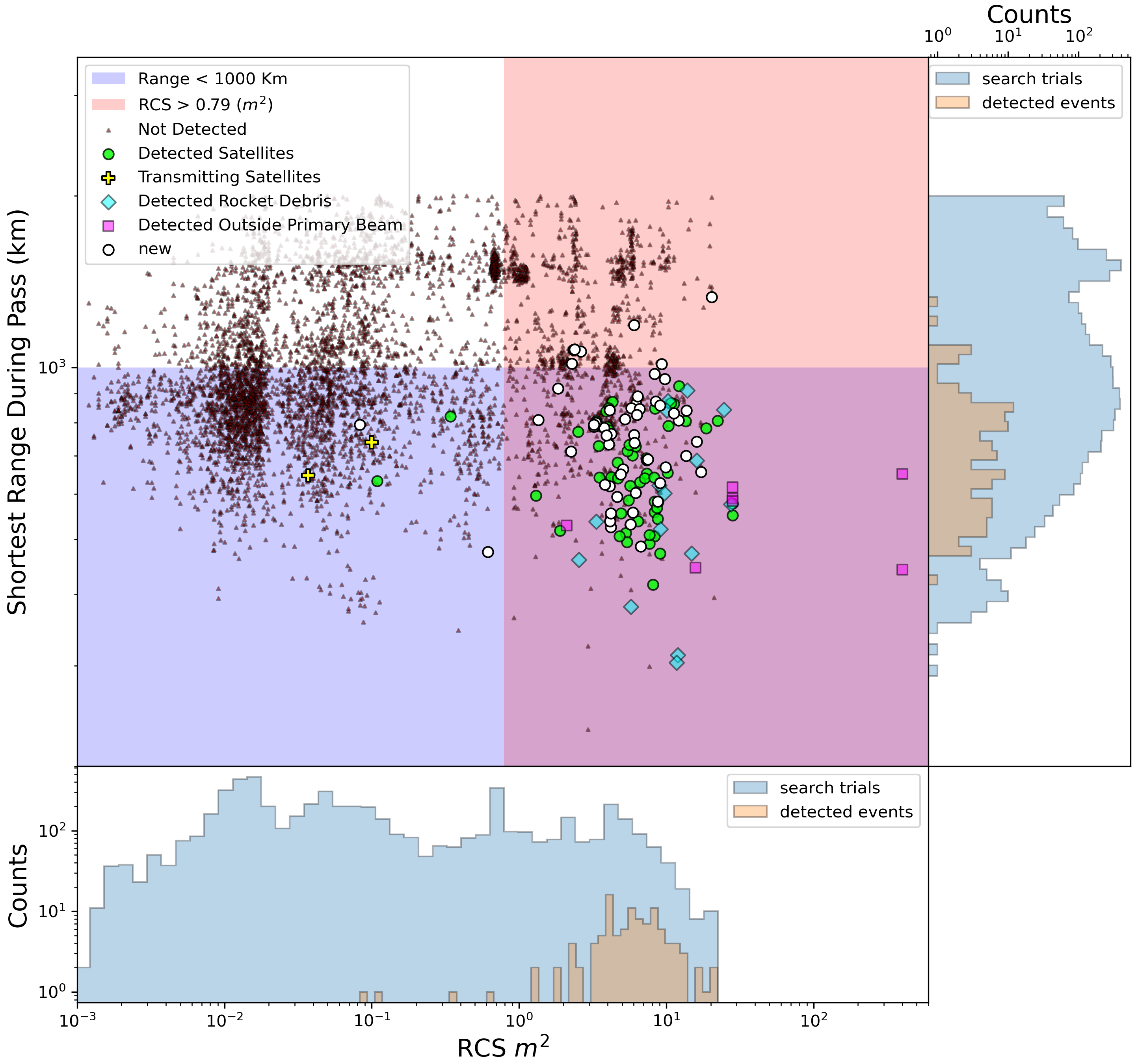}
\caption{New candidate detections (steady reflections) in range vs RCS parameter space using white circle markers. The background image is the detection summary of the blind survey performed in \cite{prabu_survey}. Note that the objects detected outside the primary beam were not considered in the histograms plotted along the right and lower axes. }
\label{Fig6RcsVsRange}
\end{center}
\end{figure*}

\begin{figure}[h!]
\begin{center}
\includegraphics[width=\linewidth,keepaspectratio]{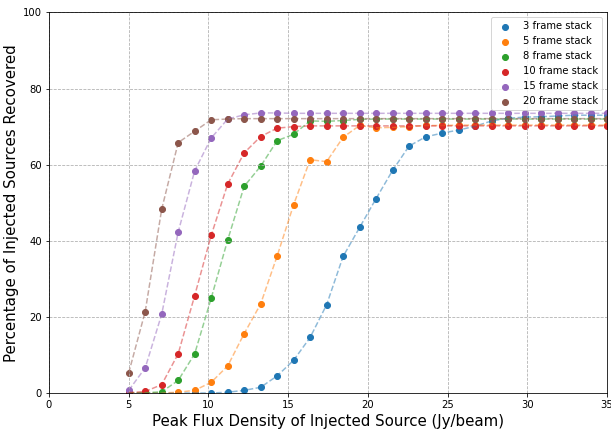}
\caption{Figure shows the percentage of injected synthetic sources recovered by our detection pipeline for different number of images/time-steps stacked. Note that the vertical scatter on the maximum recovery parentage for different number of frames stacked is within the error of the simulation. They often converged very close to the theoretical limit upon using large number of search trials, but were not performed here due to the process being computationally expensive. }
\label{FigInjectionAndRecovery}
\end{center}
\end{figure}

\begin{figure*}[h!]
\begin{center}
\includegraphics[width=\linewidth,keepaspectratio]{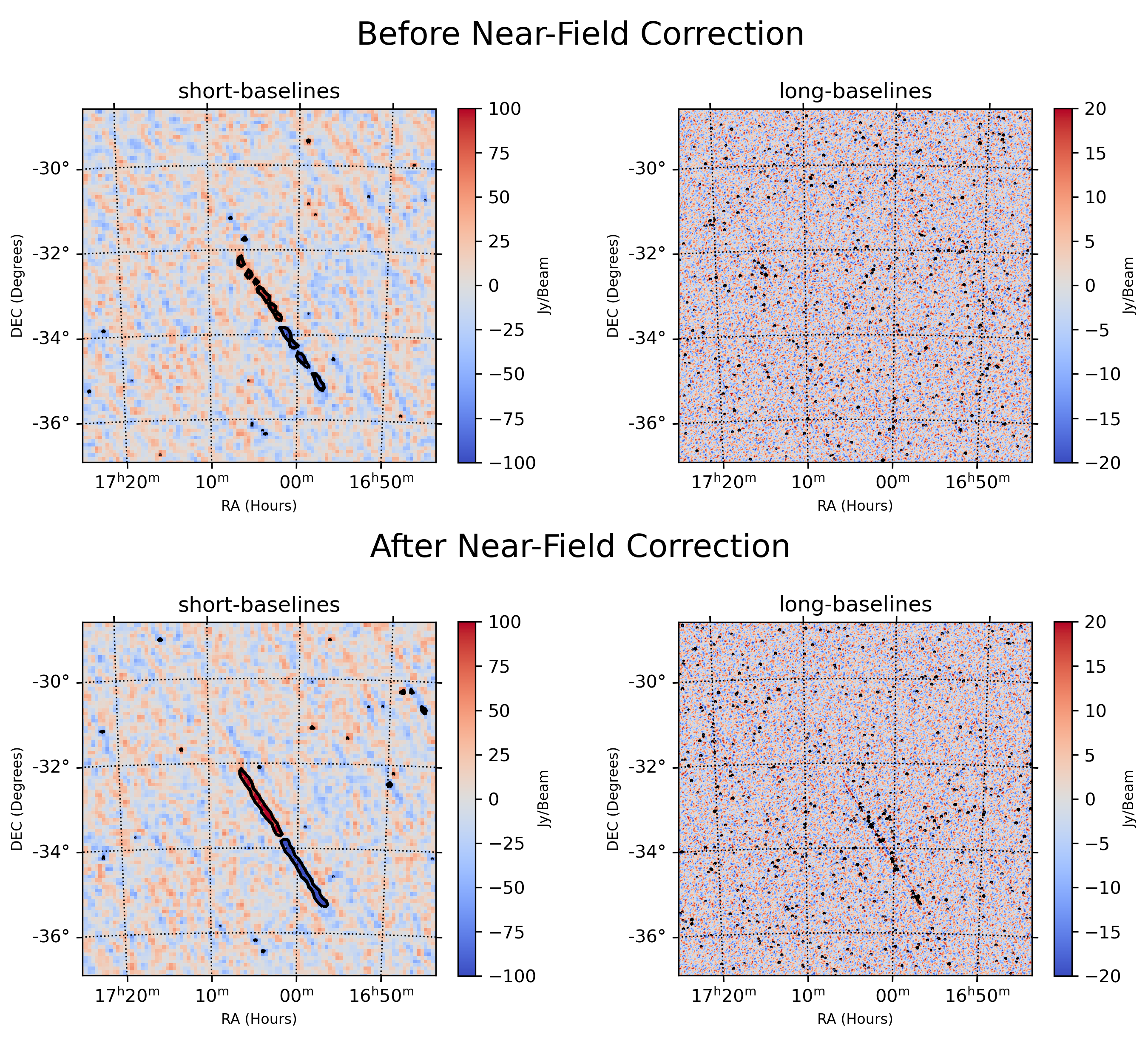}
\caption[Demonstration of the near-field imaging capability using the MWA.]{Demonstration of the near-field imaging capability with the MWA. The top two panels are the short-baseline and long-baseline difference image of a single fine channel at a given time-step, prior to applying the required phase correction. The long/short baseline cut-off used is $826$\,m, as determined by the near-field equation. In all the four panels, we use black contours to help identity bright pixels in the image (pixels whose absolute value is greater than 60 in the short baseline images and pixels whose absolute value is greater than 25 in the long baseline images). The fits files used to make the above figure can be obtained from \url{https://zenodo.org/record/5493585\#.YThmjO0zZhE}.}
\label{Fig7near-field}
\end{center}
\end{figure*}

\begin{figure*}[h!]
\begin{center}
\includegraphics[width=\linewidth,keepaspectratio]{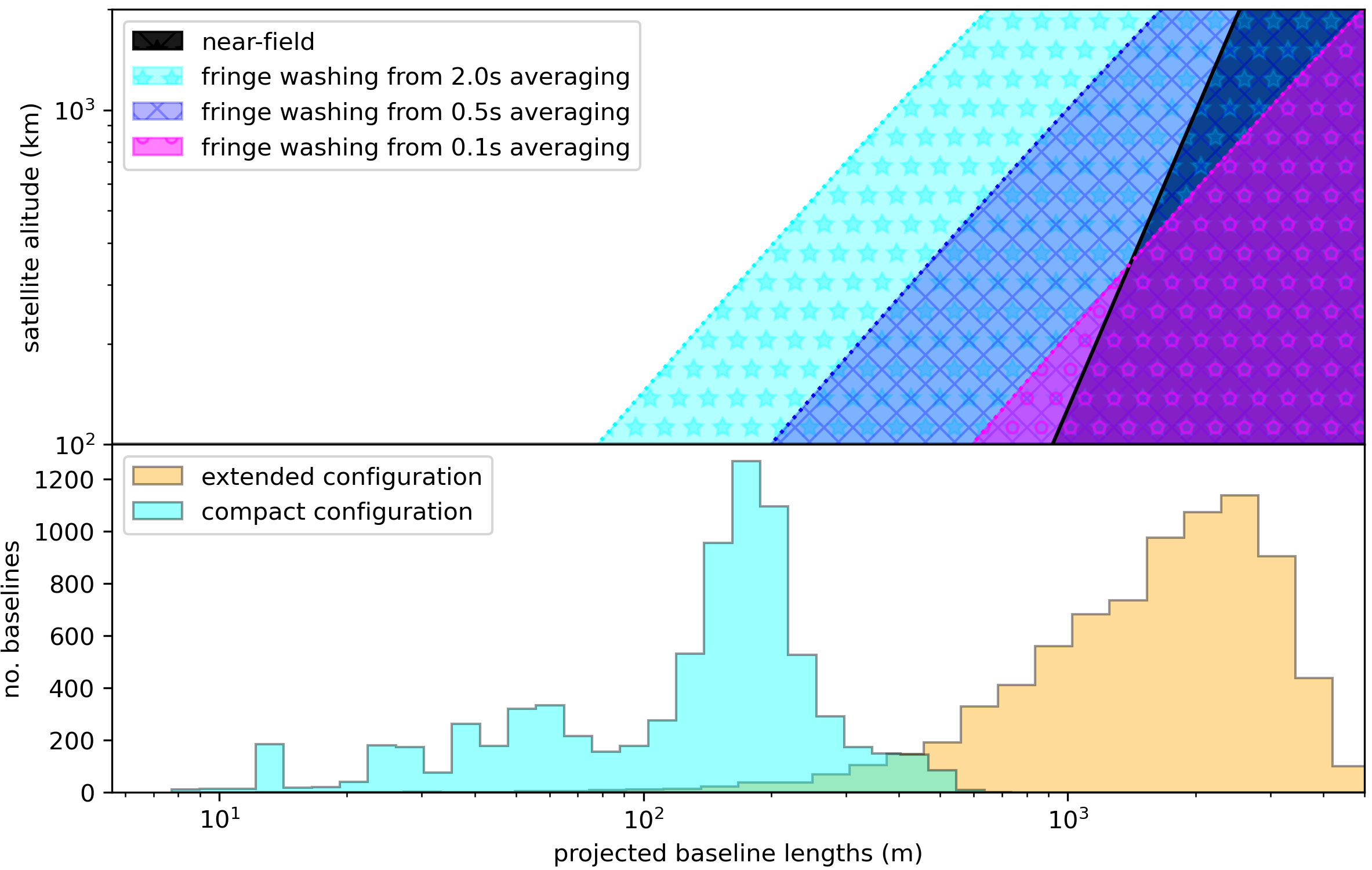}
\caption[Baselines affected by fringe-washing and near-field effects.]{Top panel shows the baselines affected by the near-field effect and fringe-washing (phase change of $\pi$) as a function of satellite altitude (angular speed). The bottom panel of the figure shows the baseline length distribution for the phase 2 compact and extended configuration.}
\label{Fig8fringe-washing}
\end{center}
\end{figure*}

The final list of new detection candidates (from Table \ref{tab1}) from the shift-stack targeted search is shown in the Radar Cross-Section\footnote{Obtained from \url{https://celestrak.com/pub/satcat.txt}} (RCS) vs range parameter space in Figure \ref{Fig6RcsVsRange}. The background image is the blind survey detection summary figure from \cite{prabu_survey}, and the new detections are annotated using white circle markers. The histograms along the right and lower axes of Figure \ref{Fig6RcsVsRange} show the distribution of all the search trials performed using shift-stack in this paper (blue) along with the total number of detections (blind survey and shift-stack in orange) obtained inside the primary beam during the observations used. The blue histogram shows the parameter space probed during the search, and the orange is what was detected (completeness of the technique) using the non-coherent methods developed in \cite{prabu_survey} and this paper.  We also show the region with RCS $>0.79$\,m$^2$ and range $<1000$\,km as the parameter space \cite{Tingay2013OnFeasibility} predicts the MWA to be sensitive towards.  The majority of the detections do fall within the predicted region. Within the predicted parameter space (bottom-right quadrant of Figure \ref{Fig6RcsVsRange}), we detect $20\%$ of the objects that passed through the primary beam. The remaining objects were likely not detected due to non-favourable transmitter-RSO-MWA reflection geometries during the pass. Also, the RCS values plotted in Figure \ref{Fig6RcsVsRange} are very likely to be much smaller in the observed FM wavelengths\footnote{the RCS measurement of the RSOs are performed using VHF/UHF/S-Band radars and are much lower in FM frequencies where the size of most RSOs are comparable to the observed wavelength}(hence we use them as a size order of magnitude guide only).

\subsubsection{Shift-Stacking Completeness}
We quantify the completeness of our shift-stacking search by performing an injection and recovery test\footnote{The code and the pool of MWA images used to perform the injection and recovery test can be obtained from \url{https://zenodo.org/record/6275009\#.YhiSc-0zZhE}}. We inject 2D Gaussian sources of varying peak flux density into real MWA images and calculate our recovery rates. Since in Section \ref{shift-stacking} we only classify events with spatially coherent signal in more than one fine channel as a detection (to remove false positives from aliased side-lobes of other bright events), we inject the synthetic source in two fine-channels and attempt to recover them. Doing so helps us also estimate the number of false-negatives in our method due to one of the injected fine frequency channels being noisier than the other (resulting in the source being recovered in only one of them). Performing the injection and recovery test using real MWA images also helps determine the impact of the instrument's thermal noise on the detection rate, and accurately captures the impact of increased confusion noise from other FM band events (such as meteor scatter, atmospheric FM ducting, and other RSOs within the FOV). Different RSOs were visible for different duration, (resulting in different number of frames stacked) within the MWA's primary beam, and hence we also test our recovery rates for different number of frames stacked.

For a given peak flux density ($S_{peak}$) and number of frames stacked ($N$), we perform $1000$ search trails (arbitrarily chosen large number) to estimate our recovery rates, and the steps involved in the injection recovery test is given below.

\begin{itemize}
    \item Step1: randomly choose two different fine channels within the FM band;
    \item Step2: retrieve $N$ difference images in the chosen fine channels;
    \item Step3: inject synthetic source in each frame of peak flux density $S_{peak}$; and
    \item Step4: stack frames and attempt recovering the synthetic source. Note the event is classified as a detection only if its is detected over $6\sigma$ in both the fine channels.
\end{itemize}

The percentages of sources recovered is shown in Figure \ref{FigInjectionAndRecovery}, and we see that we can recovery only approximately about $72$ percent of the sources. This is in agreement with our expectation\footnote{For the observations used, 376/768 fine frequency channels where within the FM band, of which 318 where unflagged. Hence, the maximum theoretical detection rate is the probability of selecting two unflagged channels from the available 376 channels, i.e, $71.5 \%$.} as about $25$ percent of the fine channels in our observations were flagged due to non-linear band-pass characteristics of the MWA's poly-phase filter bank.

\subsection{Near-Field Analysis}
The ISS passed through the primary beam of an extended array observation, and hence is used as the object of interest in this section. The motivation for using the ISS is two-fold.  First, it is a large-sized object which gives consistent reflections at multiple frequencies and secondly, it is a particularly low orbit RSO and thus is in the near-field more often than other large objects. For one of the time-steps that the ISS was detected, the MWA's baselines were divided into short and long baselines (using $826$\,m as the cut-off for the baseline separation, as determined by the near-field approximation $d<2D^{2}/\lambda$, where $\lambda$ is the wavelength, $D$ is the baseline length, and $d$ is the far-field distance), and the corresponding fine-channel difference images are shown in the top panels of Figure \ref{Fig7near-field}. We see that the streak signal is detected above $6 \sigma$ in the short baseline image while it is not detected using the long baselines.

The near-field visibility phase correction for the appropriate time-step is applied using the {\tt LEOVision} python tool. The phase-corrected difference images for the short and the long baselines are shown in the bottom panels of Figure \ref{Fig7near-field}. While the SNR of the short baseline image has increased, the previously undetected streak signal in the long baseline image now begins to appear. This demonstrates the MWA's capability to focus on the desired near-field RSO position to perform more sensitive SDA detections.

Although the signal was recovered in the long-baseline difference image after applying the phase correction, the SNR in the long-baseline difference image is much lower than the short-baseline SNR, contrary to our expectation. When the baseline cut-off was applied, $1365$ baselines were classified as short baselines and the remaining $6636$ baselines were classified as long baselines (note only $8001$ baselines were available after flagging). Because the long-baseline image has more collecting area (number of baselines), we expect it to detect the RSO signal at a higher SNR than the short baseline image, contrary to what we see in Figure \ref{Fig7near-field}. We attribute this effect to the fringe-washing of the RSO signal on the long baselines. The phase of the measured complex visibility contains information about the source position with respect to the phase-centre. The visibility averaging duration (limited to $0.5$\,s with the current MWA hardware) of the correlator is often optimised to minimise fringe-washing of celestial sources due to sidereal rotation ($0.25^{\circ}/minute$ on the equator). However, the phase of the fast moving LEO objects (e.g, ISS moves at approx. $60^{\circ}/minute$ near the zenith) considered in this work changes rapidly within the time averaging duration, thus resulting in a loss of coherence (fringe-washing) when using the long baselines. The baseline lengths affected by fringe-washing when observing an object at an altitude $alt$ can be obtained from Equation \ref{E3} (the equation is derived in Appendix \ref{appedixFringeWashing}),

\begin{equation}
    \begin{array}{l}
        
        \Delta \text{phase}  \approx 2\pi \frac{\displaystyle b }{\displaystyle \lambda \times alt} \sqrt{\frac{\displaystyle GM_{earth}}{\displaystyle R_{earth}+alt}} \Delta t
         
    \end{array}
    \label{E3}
\end{equation}

where $\Delta \text{phase}$ is the change in visibility phase, $\Delta t$ is the visibility averaging duration in $s$, $b$ is the length (in $m$) of the baseline component parallel to the satellite motion, $\lambda$ is the wavelength in $m$, G ($m^{3}kg^{-1}s^{-2}$) is the gravitational constant, $M_{earth}$ ($kg$) is the mass of Earth, and $R_{earth}$ ($m$) is the radius of Earth. Using Equation \ref{E3}, we show the baseline lengths (parallel to the motion of the satellite, and the parallel component of baselines of other orientations) affected by the fringe-washing effect (visibility phase change of $\pi$ radians) in the top panel of Figure \ref{Fig8fringe-washing} (for $2.0$\,s, $0.5$\,s, and $0.1$\,s visibility averaging). Note that the figure also shows the baseline lengths affected by the near-field effect (independent of the visibility averaging duration).  

\begin{figure}[h!]
\begin{center}
\includegraphics[width=\linewidth,keepaspectratio]{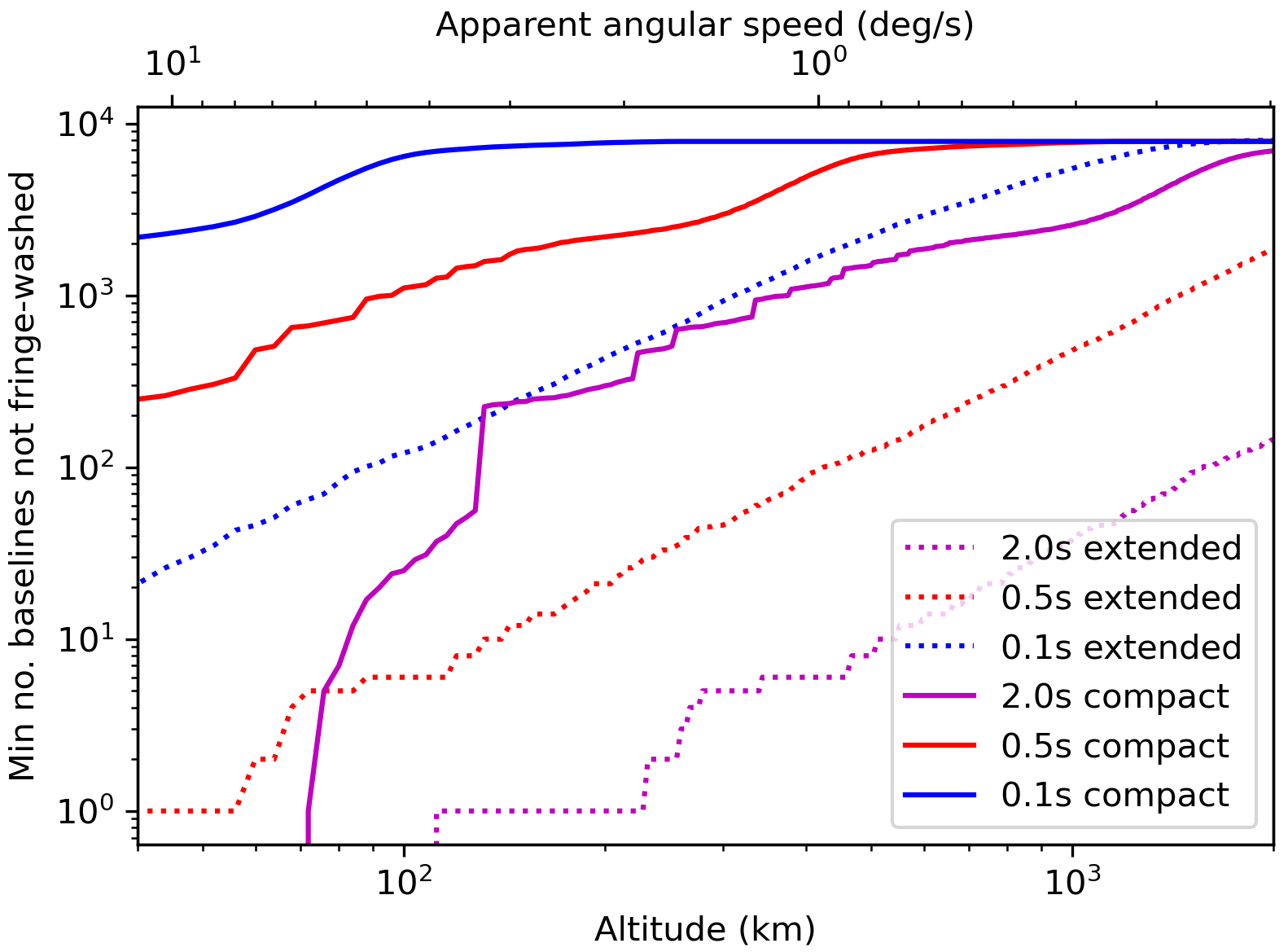}
\caption{The minimum number of baselines not fringe-washed as a function of satellite altitude (or angular speed). The actual number of baselines not affected depends on the direction of the satellite pass (as baselines perpendicular to the satellite motion are not fringe-washed). }
\label{Fig9noBaselines}
\end{center}
\end{figure}

The bottom panel of Figure \ref{Fig8fringe-washing} also shows the baseline distribution of the MWA Phase 2 extended array and the compact array. As we go from short baselines towards the longer baselines, from Figure \ref{Fig8fringe-washing}, we see that the fringe-washing comes into effect much before the near-field de-focusing effect (plotted using equation $d<2D^{2}/\lambda$) for the $2$\,s time-averaging used. Hence, in Figure \ref{Fig7near-field} we cannot recover the RSO signal with the longer baseline difference image due to fringe-washing. However, with the Phase 3 MWA, we will be able to sample visibilities with $0.1$\,s time-averaging and more sensitive SDA observations with longer baselines should be possible. The minimum number of baselines not fringe-washed\footnote{since we ignore the baseline orientation with respect to the direction of satellite pass, the number of baselines not fringe-washed in the plot can be treated as the worst case scenario. In reality, the only the baselines with a component parallel to the satellite pass will be fringe-washed and hence the figure can be used as a limit on the minimum number of baselines not fringe-washed.} as a function of altitude is shown in Figure \ref{Fig9noBaselines}. We see that the sensitivity of the extended array with the new Phase 3 correlator ($0.1$\,s averaging) is more sensitive (as more baselines are not fringe-washed) than the current hex configuration blind survey performed in \cite{prabu_survey} (compact configuration with $2$\,s averaging).

\section{CONCLUSION}
\label{conclusion}
In this paper we have developed and demonstrated two different methods to improve the detection sensitivity of the non-coherent MWA SDA system. The first RSO search method is called shift-stacking. The method  increases the SNR of faint RSO signals by averaging the signal along the predicted trajectory, through coherent stacking. We test the method for all the objects that passed through the MWA's primary beam during the $20$ hours of observation used during our previous blind survey \citep{prabu_survey}. The shift-stacking targeted search resulted in $55$ new detections that were previously not detected by our blind detection pipeline, demonstrating that the shift-stacking method is able to probe a weaker population of signals as it uses prior information about the object's pass to perform signal stacking. The MWA shift-stacking pipeline also identifies the in-track and cross-track offsets of the RSO detections, and these measurements can be used to perform RSO catalog updates.

The second method performs detections of weak RSO signals by re-focusing the interferometer to the desired near-field RSO location. It does so by calculating the apparent delay as seen by a baseline and converting it into a visibility phase correction. This method was proved to work effectively using an extended array observation of the ISS. The previously undetected ISS FM reflection signal in the long-baseline difference image of the observation was recovered after applying the required near-field phase correction. However, the recovered signal was weaker than expected, due to visibility fringe-washing. The phase of the ISS signal changed rapidly (during the correlator integration time of $0.5$\,s) resulting in de-correlation of the signal. However, with the ongoing upgrade of MWA to Phase 3, we should be able to perform more sensitive near-field detections due to being able to sample the apparent sky with $0.1$\,s averaging (reduces fringe-washing).

In summary, the implementation of the shift-stack method has increased the detection rate of the non-coherent MWA SDA system by $75\%$. Any future SDA dedicated arrays that may be built using MWA-like technology should incorporate short baselines to avoid fringe-washing of RSO signals or incorporate correlators that allow smaller integration times (higher data rate), due to the issues discussed in this paper. In the future, the near-field phase correction software developed can be modified to determine the 3D location of the target by iterative maximisation of the RSO signal, by varying its predicted location (the SNR of the focused signal is maximum when the predicted near-field location matches with its true position).

\begin{acknowledgements}
This scientific work makes use of the Murchison Radio-astronomy Observatory, operated by CSIRO. We acknowledge the Wajarri Yamatji people as the traditional owners of the Observatory site. Support for the operation of the MWA is provided by the Australian Government (NCRIS), under a contract to Curtin University administered by Astronomy Australia Limited. We acknowledge the Pawsey Supercomputing Centre which is supported by the Western Australian and Australian Governments. Steve Prabu would like to thank Innovation Central Perth, a collaboration of Cisco, Curtin University,  Woodside and CSIRO’s Data61, for their
scholarship.

\subsection*{Sofware}
We acknowledge the work and the support of the developers of the following Python packages:
Astropy \citep{theastropycollaboration_astropy_2013,astropycollaboration_astropy_2018}, Numpy \citep{vanderwalt_numpy_2011}, Scipy  \citep{jones_scipy_2001}, matplotlib \citep{Hunter:2007}, seaborn\footnote{\url{https://seaborn.pydata.org/index.html}} python-casacore \footnote{\url{https://github.com/casacore/python-casacore}},and SkyField\footnote{\url{https://rhodesmill.org/skyfield/}}. The work also used WSCLEAN \citep{offringa-wsclean-2014,offringa-wsclean-2017} for making fits images and DS9\footnote{\href{http://ds9.si.edu/site/Home.html}{ds9.si.edu/site/Home.html}} for visualization purposes. Adobe Illustrator was used for creating Figure \ref{Fig2near-field}.

\end{acknowledgements}

\begin{appendix}

\section{Visibility fringe-washing as a function of satellite altitude}
\label{appedixFringeWashing}
The classical mechanics equation for the linear velocity of a satellite in a circular orbit around Earth is given by

\begin{equation}
    \begin{array}{l}
        
    v = \sqrt{\frac{\displaystyle GM}{\displaystyle R + alt}}

    \end{array}
    \label{EA.1}
\end{equation}

where $v$ ($m/s$) is the linear velocity of the satellite, $G$ is the gravitational constant ($6.67408 \times 10^{-11} m^{3} kg^{-1} s^{-2}$), $R$ ($6.371\times10^{6}$\,m) is the radius of Earth, $M$ is the mass of Earth ($5.972\times 10^{24}$\,kg) and $alt$ is the altitude of the considered satellite.

The fringe-rate (the time rate of change of the visibility phase) as measured by a zenith pointed East-West baseline is given by (obtained from \cite{marr2015fundamentals})

\begin{equation}
    \begin{array}{l}
        
    \frac{\displaystyle \delta \Phi}{\displaystyle \delta t} \approx 2\pi \omega \frac{\displaystyle b}{\displaystyle \lambda}
    \end{array}
    \label{EA.2}
\end{equation}

and for small integration times Equation \ref{EA.2} becomes

\begin{equation}
    \begin{array}{l}
        
    \frac{\displaystyle \Delta \Phi}{\displaystyle \Delta t} \approx 2\pi \omega \frac{\displaystyle b}{\displaystyle \lambda} \\
    
    \end{array}
    \label{EA.3}
\end{equation}

\begin{equation}
    \begin{array}{l}
    \Delta \Phi \approx 2\pi \omega \frac{\displaystyle b}{\displaystyle \lambda} \Delta t
       \end{array}
    \label{EA.3b}
\end{equation}

For an observer on the surface (i.e, the MWA), the apparent angular motion (for small angles near the zenith) can be derived using the derivative form of $s=r\theta$ relating angular velocity with linear velocity ($v = r \omega$). The apparent angular velocity $\omega$ for a satellite (near the zenith) can be determined using Equation \ref{EA.1}

\begin{equation}
    \begin{array}{l}
    \omega = \frac{\displaystyle v }{\displaystyle radius} = \frac{\displaystyle v}{\displaystyle alt} \\ 
     \implies \quad \omega = \frac{\displaystyle 1 }{\displaystyle alt} \sqrt{\frac{\displaystyle GM}{\displaystyle R + alt}}
         
    \end{array}
    \label{EA.4}
\end{equation}

Substituting the value of $\omega$ from Equation \ref{EA.4} in Equation \ref{EA.3}, we get the function relating the change in visibility phase as measured by a baseline for the apparent motion of a satellite.

\begin{equation}
    \begin{array}{l}
    
    \Delta \Phi \approx 2\pi \frac{\displaystyle b }{\displaystyle \lambda \times alt} \sqrt{\frac{\displaystyle GM}{\displaystyle R + alt}}  \Delta t

    \end{array}
    \label{EA.5}
\end{equation}

\section{List of new detections}
\label{listofdetections}

\begin{table*}[h!]
\caption{All the new events detected by the shift-stacking targeted search. We list the North American Aerospace Defence  (NORAD) catalog number of all the objects detected along with its RCS, shortest range during observation, and its apparent peak flux density.}
\label{tab1}
\centering
\begin{tabular*}{\textwidth}{c c c c c c}
Observation & Name  & Norad ID & RCS & Shortest Range & Apparent Peak Flux Density\\
ID &  &   & $m^{2}$ & km & (Jy/beam)  \\
\hline \hline

1165782736 & DELTA 2 R/B(1) & 24809 & 9.88 & 668 & 5.41 \\
1165782496 & DELTA 2 R/B(1) & 23640 & 9.75 & 954 & 5.65 \\
1165776496 & CZ-2D R/B & 36597 & 8.72 & 582 & 6.09 \\
1165776256 & SL-24 R/B & 31123 & 5.23 & 811 & 7.85 \\
1165775896 & FENGYUN 3B & 37214 & 6.21 & 863 & 7.43 \\
1165771576 & SUOMI NPP & 37849 & 5.78 & 848 & 8.78 \\
1165770376 & IRIDIUM 43 & 25039 & 3.20 & 788 & 5.18 \\
1165770256 & ATLAS AGENA D R/B & 2144 & 6.04 & 761 & 5.32 \\
1165768576 & ASTRO-H (HITOMI) & 41337 & 6.16 & 603 & 7.11 \\
1165767376 & OCEANSAT-2 & 35931 & 4.06 & 732 & 6.14 \\
1165766296 & SL-16 R/B & 24298 & 8.49 & 871 & 6.79 \\
1165765936 & OAO 1 & 2142 & 12.1 & 807 & 4.33 \\
1165765816 & PSLV R/B & 25759 & 6.06 & 737 & 5.20 \\
1165763536 & DELTA 2 DEB [DPAF] & 29110 & 5.07 & 662 & 4.61 \\
1165763536 & KAZEOSAT 1 & 39731 & 4.22 & 765 & 3.66 \\
1165760896 & COSMOS 2486 & 39177 & 16.0 & 741 & 2.18 \\
1165759696 & KORONAS-FOTON & 33504 & 4.18 & 524 & 6.90 \\
1165757896 & KMS 4 & 41332 & 0.61 & 475 & 3.29 \\
1165757056 & CZ-2C DEB & 40288 & 0.08 & 793 & 2.86 \\
1165753576 & CAMEO and DELTA 1 R/B & 11081 & 8.30 & 975 & 6.23 \\
1160505712 & IRIDIUM 7 & 24793 & 3.32 & 799 & 5.72 \\
1160505112 & IRIDIUM 6 & 24794 & 3.23 & 794 & 8.27 \\
1160504872 & SL-24 DEB & 33318 & 4.64 & 593 & 6.12 \\
1160504272 & SL-16 R/B & 17974 & 9.05 & 858 & 10.0 \\
1160503672 & IRS-1D & 24971 & 3.91 & 760 & 9.29 \\
1160502232 & FORTE & 24920 & 1.35 & 809 & 6.57 \\
1160500192 & SL-14 R/B & 10974 & 4.12 & 619 & 4.62 \\
1160499712 & SUZAKU (ASTRO-EII) & 28773 & 4.15 & 538 & 7.94 \\
1160499352 & INTERCOSMOS 25 & 21819 & 7.32 & 688 & 8.92 \\
1160498872 & CZ-2C R/B & 40262 & 7.47 & 690 & 10.5 \\
1160489992 & PSLV R/B & 41620 & 6.71 & 485 & 9.66 \\
1160489152 & METOP-B & 38771 & 13.6 & 840 & 6.87 \\
1160489032 & H-2A R/B & 38341 & 17.2 & 655 & 6.57 \\
1160483992 & COSMOS 1833 & 17589 & 4.11 & 841 & 7.97 \\
1160482912 & FENGYUN 3A & 32958 & 6.58 & 845 & 9.37 \\
1160479912 & COSMOS 1300 & 12785 & 5.69 & 531 & 16.2 \\
1157489632 & SL-24 R/B & 31699 & 5.92 & 557 & 21.3 \\
1157469232 & KORONAS-FOTON & 33504 & 4.18 & 554 & 17.9 \\
1157459032 & ALOS (DAICHI) & 28931 & 13.6 & 700 & 33.5 \\
1157406472 & SL-14 R/B & 14820 & 3.82 & 623 & 4.42 \\
1157406472 & TSX-5 & 26374 & 1.84 & 919 & 3.42 \\
1157401672 & SL-3 R/B & 7275 & 6.37 & 889 & 6.34 \\
1157398072 & BREEZE-M DEB [TANK] & 36594 & 6.29 & 825 & 4.05 \\
1157384272 & SJ-11-07 & 40261 & 2.26 & 712 & 8.18 \\
1157378272 & METOP-A & 29499 & 11.2 & 830 & 8.86 \\
1165769776 & COSMOS 860 & 9486 & 2.28 & 1015 & 4.71 \\
1160496712 & SL-12 R/B(2) & 27473 & 20.3 & 1329 & 3.73 \\
1160495392 & YAOGAN 25A & 40338 & 2.33 & 1074 & 5.19 \\
\hline\hline
\end{tabular*}
\tabnote{Continued on next page...}
\end{table*}

\addtocounter{table}{-1}
\begin{table*}[h!]
\caption{...continued from previous page.}
\centering
\begin{tabular*}{\textwidth}{c c c c c c}
Observation & Name  & Norad ID & RCS & Shortest Range & Apparent Peak Flux Density\\
ID &  &   & $m^{2}$ & km & (Jy/beam)  \\
\hline \hline
1160495392 & YAOGAN 25C & 40340 & 2.63 & 1066 & 6.21 \\
1160495392 & YAOGAN 25B & 40339 & 2.39 & 1075 & 3.09 \\
1157484832 & CZ-3 R/B & 20474 & 9.30 & 1014 & 26.3 \\
1165762936 & CBERS 4 & 40336 & 3.78 & 783 & 6.94 \\
1160498512 & SL-8 R/B & 11170 & 6.04 & 1187 & 5.08 \\
1160491312 & SL-24 DEB & 35689 & 4.88 & 649 & 8.64 \\
1160492992 & DELTA 1 R/B(1) & 10793 & 9.06 & 627 & 10.8 \\
\hline \hline
\end{tabular*}

\end{table*}

\newpage

\end{appendix}



\clearpage
\printbibliography

\end{document}